\documentclass[doublecol,linenumbers]{epl2} 

\title{Reply to comment by M. Pitkin on ``Measurements of Newton's gravitational constant and the length of day"}
\shorttitle{Reply to comment by M. Pitkin} 

\author{J. D. Anderson\inst{1\footnote{Retired.}} \and G. Schubert\inst{2} \and V. Trimble\inst{3} \and M. R. Feldman\inst{4}}
\shortauthor{J. D. Anderson \etal}

\institute{                    
  \inst{1} Jet Propulsion Laboratory, California Institute of Technology - Pasadena, CA 91109, USA \\
  \inst{2} Department of Earth, Planetary and Space Sciences, University of California, Los Angeles \\ Los Angeles, CA 90095, USA \\
  \inst{3} Department of Physics and Astronomy, University of California Irvine - Irvine CA 92697, USA \\
  \inst{4} Private researcher - Los Angeles, CA 90046, USA
}

\begin{document}

\maketitle

\begin{figure}
\includegraphics[width=8.0cm]{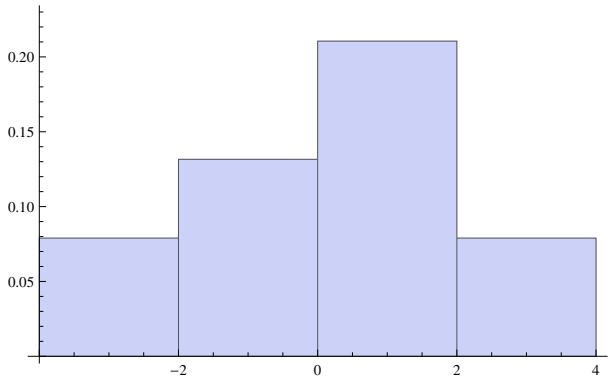}
\caption{Probability density function of the normalized residuals about a constant plus two-period sinusoid model (hypothesis 3).}
\label{PlotPDFTwoPeriod}
\end{figure}

\begin{figure}
\includegraphics[width=8.0cm]{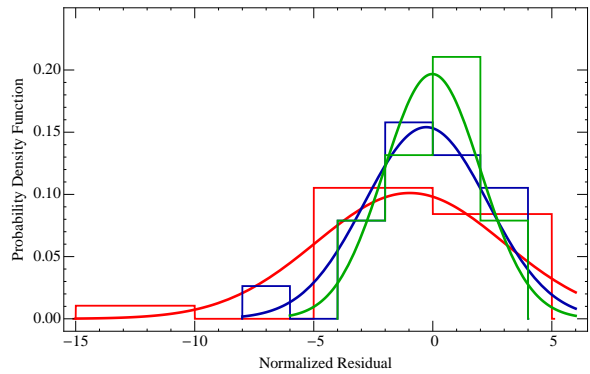}
\caption{Probability density functions of the normalized residuals for all three hypotheses (red: hypothesis 1, blue: hypothesis 2, green: hypothesis 3). Overlayed are fitted normal distributions for each hypothesis with clear indication that hypothesis 3 residuals have a tighter Gaussian fit than the others.}
\label{PlotPDFNormalOverlay}
\end{figure}

The comment by M. Pitkin \cite{Pitkin2015} on our {\it EPL} article ``Measurements of Newton's gravitational constant and the length of day" claims to provide evidence that a constant $G$ measurement model with an additional Gaussian noise term is ``hugely favoured" over models employing sinusoidal terms when using a Bayesian model selection procedure. Unfortunately, we were unable to replicate his claims with our own independent analysis testing the hypotheses of the following three scenarios for the $G$ measurements:
\begin{enumerate}
\item Constant value: $G = a_0$.
\item Constant plus a sinusoidal term with period of approximately $6$ years: $G = a_0 + a_1\cos\bigg(\frac{2\pi t}{P_1}\bigg) + b_1 \sin\bigg(\frac{2\pi t}{P_1}\bigg)$.
\item Constant plus sinusoidal terms with two different periods of approximately $6$ years and $1$ year: $G = a_0 + a_1\cos\bigg(\frac{2\pi t}{P_1}\bigg) + b_1 \sin\bigg(\frac{2\pi t}{P_1}\bigg) + a_2\cos\bigg(\frac{2\pi t}{P_2}\bigg) + b_2 \sin\bigg(\frac{2\pi t}{P_2}\bigg)$.
\end{enumerate}
\begin{figure}
\includegraphics[width=8.0cm]{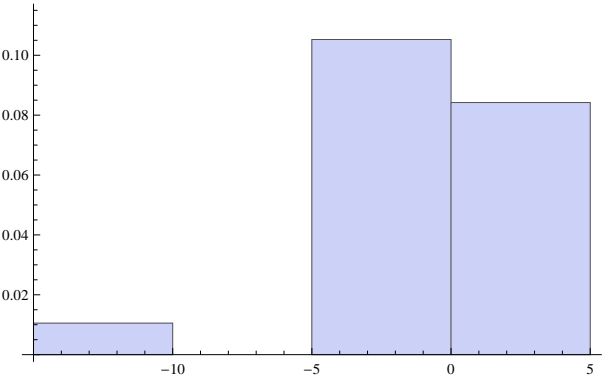}
\caption{Probability density function of the normalized residuals about a weighted mean $G$ model (hypothesis 1).}
\label{PlotPDFMean}
\end{figure}
We used a non-linear regression analysis with a minimization of the L1 norm to determine the best fit values for the input parameters of each of the above cases (i.e. for $a_i, b_i$ and $P_i$ values). After fitting to the $G$ data, we found normalized $\sigma$ values of $4.0$, $2.6$ and $2.0$ for the weighted residuals of scenarios $1$, $2$ and $3$, respectively, suggesting the two-period model is favored. We also computed histograms for the 19 weighted residuals and best fit probability density functions for the three hypotheses. Importantly, the probability density function of the weighted residuals about a mean value of the $G$ measurements (hypothesis 1) follows more of a uniform distribution whereas for the two-period sinusoidal model (hypothesis 3) the probability density function of the weighted residuals appears to follow a normal distribution, suggesting a possible error in Pitkin's analysis. See Figs.~\ref{PlotPDFMean}-\ref{PlotPDFNormalOverlay} below for our outputs from {\tt Mathematica}. Thus, we stand by our conclusions of potential periodic terms in the reported $G$ measurements (see our added appendix of \cite{Anderson2015} in response to \cite{Schlamminger2015} for our logic with a two-period sinusoidal model).

\end{document}